\documentclass{jfm}

\usepackage{graphicx}
\usepackage{newtxtext}
\usepackage{newtxmath}
\usepackage{natbib}
\newcommand{\mockalph}[1]{}
\usepackage{hyperref}
\hypersetup{
    colorlinks = true,
    urlcolor   = blue,
    citecolor  = blue,
    linkcolor   = blue,
}

\title[Parabolic velocity profile causes drift of inertial prolate spheroids]{Parabolic velocity profile causes drift of inertial prolate spheroids -- but gravity is stronger}

\author[J. Bagge, T. Ros\'{e}n, F. Lundell, A.-K. Tornberg]%
{J. Bagge\aff{1},
T. Ros\'{e}n\aff{2,3},
F. Lundell\aff{2,3}
and A.-K. Tornberg\aff{1} \corresp{\email{akto@kth.se}}}

\affiliation{\aff{1}KTH Mathematics, Linn\'{e} FLOW Centre/Swedish e-Science Research Centre, Royal Institute of Technology, SE-100 44 Stockholm, Sweden
\aff{2}KTH Mechanics, Linn\'{e} FLOW Centre, Royal Institute of Technology, SE-100 44 Stockholm, Sweden
\aff{3}Wallenberg Wood Science Center, Royal Institute of Technology, SE-100 44 Stockholm, Sweden}

\begin{document}
\maketitle

\begin{abstract}
  Motion of elongated particles in shear is studied. In applications where particles are much heavier than the carrying fluid, e.g. aerosols, the influence of particle inertia dominates the particle dynamics. Assuming that the particle only experiences a local linear velocity profile, its rotational and translational motion are independent. However, we show that quadratic terms of the local velocity profile combined with particle inertia cause a lateral drift of prolate spheroidal particles. We find that this drift is maximal when particle inertial forces are of the same order of magnitude as viscous forces, and that both extremely light and extremely heavy particles have negligible drift. In the non-inertial case, the particle rotates according to the local linear velocity profile, with each instantaneous orientation corresponding to a velocity that gives zero force on the particle. This results in a translational motion in the flow direction with periodic velocity fluctuations. With added particle inertia, the particle is slow to react to the surrounding fluid motion and the particle will switch between slower and faster rotation compared to the zero-force solution. The final motion that gives zero integrated force over a rotational period, is a motion with a lateral drift. We show that this drift is purely an effect of the non-sphericity of the particle and its \emph{translational} inertia, while \emph{rotational} inertia is negligible. Finally, although this inertial drift will contribute to the lateral motion of heavy elongated particles in channel flow, sedimentation due to gravity will dominate in any practical application on earth.
\end{abstract}

\begin{keywords}
Suspension flow, particle inertia, quadratic flow
\end{keywords}

\section{Introduction}\label{sec:Intro}

There are numerous practical examples where the understanding of the motion of heavy particles suspended in lighter fluid is important. Typically, this will apply to any solid particle in air, commonly called \emph{aerosols}. In the atmosphere, aerosols scatter light and thus also affect the global radiation budget \citep{Hollander}. These particles also serve as condensation nuclei for cloud formation and their motion inside the clouds is important for understanding rain initiation \citep{Falkovich01,Falkovich02} as well as snow crystal growth \citep{GavzePinsky}. 

Closer to earth, aerosols are typically associated with vehicle and industrial emissions causing severe health problems \citep{Morawska02}. Also asbestos fibers in isolation materials in buildings can be directly linked to formation of lung cancer \citep{Miserocchi}. When controlling and separating these particles from the suspending fluid, as well as understanding the deposition in the airways, it is important to find how these particles behave in channel flows. Especially important is to understand any physical process causing lateral migration of particles as this will determine if particles for example are concentrated in the middle of a channel or towards the walls. 

Suspension flows are commonly modelled with spherical particles, due to the many models available to calculate the motion \citep{Crowe}. For spherical particles it is known that inertia of the surrounding fluid can cause three types of lateral motion of the spherical particles. Firstly, if the particle is not moving with the same velocity as the surrounding fluid, i.e.~there is a \emph{slip velocity}, and the particle is rotating in a constant flow without gradients, there is a lateral force on the particle called a \emph{Magnus} force \citep{Crowe}. Secondly, if the particle has a slip velocity and is rotating in a linear shear flow, there is a lateral force on the particle called a \emph{Saffman} force \citep{Crowe}. Thirdly, if the particle is suspended in a regular pipe flow, the parabolic velocity profile causes itself a migration of particles away from the centreline. As a particle move closer to the wall, the lateral migration force is balanced by pressure that is built up between the particle and the wall, causing the particle to find an equilibrium radial position in the channel. This effect is referred to as the \emph{Segr\'{e}-Silberberg effect} \citep{SegreSilberberg}. All these effects are however only present when fluid inertia is relevant. If fluid inertia is negligible, which typically is true for aerosols, there is no lateral drift of spherical particles either caused by a slip velocity or a quadratic velocity profile. 

When it comes to modelling non-spherical particles, it is common to study ellipsoids, and particularly ellipsoids with rotational symmetry called \emph{spheroids}. In the absence of fluid inertia, \citet{lamb1932hydrodynamics} provided analytical expressions of the force on an ellipsoid in a constant flow given the orientation of the particle and \citet{Jeffery} provided analytical expressions of the torque on an ellipsoid in a linear flow field. The force on a suspended inertial ellipsoid can thus be determined by the instantaneous slip velocity and the torque can be determined by local velocity gradients. The results by \citet{Jeffery} has however mainly been used to study particles that are not affected by particle inertia and therefore will assume a rotation that gives zero torque. \citet{LundellCarlsson} coupled the torques by \citet{Jeffery} to the equations of motion of the particle, and thus the rotational motion of an inertial particle could be simulated assuming that the surrounding flow still had zero inertia. This approach has been used in several studies using Lagrangian particle tracking (LPT) methods in e.g.~turbulent channel flow with included particle inertia \citep[e.g.][]{Zhao}.

For the sake of discussion, we will now consider a plane laminar Poiseuille flow with a quadratic velocity profile with suspended spheroidal particles and a dilute concentration. If we would be using a LPT method for determining the particle dynamics that uses the gradient of the velocity but no higher derivatives, we would find that the particles are nicely following the straight streamlines and rotate according to the local shear given by the distance from the centreline. The orientational dynamics of the particles will then be fully determined by their behaviour in a simple shear flow. We can add both fluid and particle inertia to the rotation owing to the recent effort in mapping out the orientational behaviour of spheroids in simple shear flow \citep{Rosen_TZ}. Due to the symmetry of the simple shear flow, there is no lateral drift of the spheroidal particles. Any drift must thus be caused by higher order derivatives of the local velocity field. \citet{Chwang} studied the spheroidal particle without either particle or fluid inertia in a quadratic flow and found no lateral drift. It is still likely that inertia of the surrounding fluid will induce a Segr\'{e}-Silberberg effect also for the spheroidal particles due to the parabolic velocity profile. In this work we will show that the influence of particle inertia combined with a quadratic velocity profile will cause a lateral drift even when fluid inertia is neglected. The present results thus provide new fundamental knowledge about the migration of aerosols in channel flows.

The flow problem is defined in Section~\ref{sec:FlowProb} and the
numerical method used is described in Section~\ref{sec:Method}.
The results are presented in Section~\ref{sec:Results} and
discussed in Section~\ref{sec:Discussion}. Two important aspects:
consequences for LPT simulations and the effect of gravity, are
investigated in Sections~\ref{sec:LPTcons}~and~\ref{sec:GravForceEffects}, respectively. Finally the conclusions are summarized in Section~\ref{sec:Conclusions}.

\section{Flow problem}\label{sec:FlowProb}

\begin{figure}
\begin{center}
\includegraphics{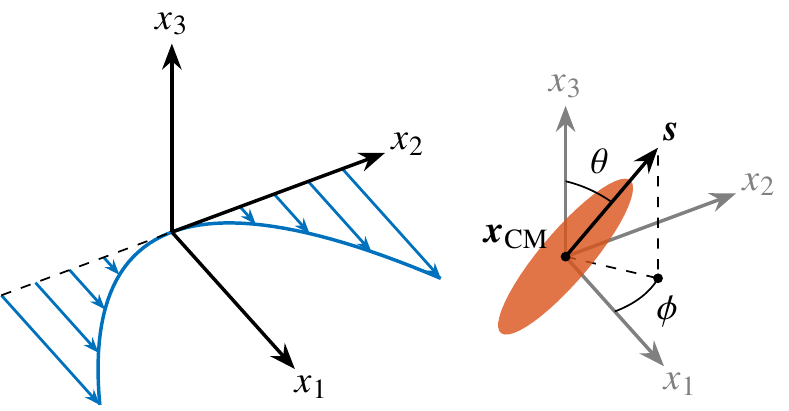}
\caption{\label{fig:FlowProb}Illustration of the flow problem; a prolate spheroidal particle is suspended in a quadratic velocity profile; the position of the particle is given by the centre-of-mass $\boldsymbol{x}_\text{CM}$ and the orientation is given by the symmetry axis $\boldsymbol{s}$ and the Euler angles $\theta$ and $\phi$.}
\end{center}
\end{figure}

A prolate spheroid with major semi-axis $l$ is suspended in a quadratic background flow according to figure~\ref{fig:FlowProb}. The unit vectors $\boldsymbol{e}_1$, $\boldsymbol{e}_2$ and $\boldsymbol{e}_3$ denote the flow direction, the velocity gradient direction and the vorticity direction, respectively. The spatial coordinates are given in dimensional form as $\boldsymbol{x}^*=(x^*_1,x^*_2,x^*_3)$. In this work, we will mainly use the non-dimensional coordinates scaled by the particle major semi-axis, i.e.~$\boldsymbol{x}=\boldsymbol{x}^*/l=(x_1,x_2,x_3)$. The non-dimensional coordinates of the particle centre-of-mass is denoted by $\boldsymbol{x}_\text{CM}$.

The prolate spheroidal particle with major semi-axis $l$ and equatorial radius $l/r_\text{p}$, where $r_\text{p}$ is the particle aspect ratio (length/width), is described through
\begin{equation}
  \label{eq:Spheroid}
  x'^{2}_1+r_\text{p}^2x'^{2}_2+r_\text{p}^2x'^{2}_3=1.
\end{equation}
The non-dimensional coordinates $x'_1, x'_2$ and $x'_3$ are scaled by the particle length $l$ and refer to the body-fixed system spanned by unit vectors $\boldsymbol{e}'_1, \boldsymbol{e}'_2$ and $\boldsymbol{e}'_3$. The orientation of the particle is given by the direction of the unit vector along the symmetry axis $\boldsymbol{s}$ (note that $\boldsymbol{s}=\boldsymbol{e}'_1$). We also express the orientation using the spherical coordinate angles $\theta$ and $\phi$ such that $\boldsymbol{s}=(s_1,s_2,s_3)=(\sin\theta\cos\phi,\sin\theta\sin\phi,\cos\theta)$ according to fig.~\ref{fig:FlowProb}. 

The (dimensional) background flow is given by $\boldsymbol{u}^*_\text{bg}(\boldsymbol{x})=(1/2)\dot{\gamma}'l^2x_2^2\boldsymbol{e}_1$, where $(1/2)\dot{\gamma}'$ is the curvature of the velocity profile. The local shear rate at the particle position thus becomes $\dot{\gamma}_\text{L}(x_\text{CM,2}^*)=\dot{\gamma}'|x_\text{CM,2}^*|$. As a global time scale in this flow problem, we choose $\dot{\gamma}_\text{G}^{-1}=[\dot{\gamma}_\text{L}(x_\text{CM,2}^*=l)]^{-1}=(\dot{\gamma}'l)^{-1}$. With these spatial and temporal scalings, the non-dimensional form of the background flow becomes
\begin{equation}
\boldsymbol{u}_\text{bg}(\boldsymbol{x})=\frac{1}{2}x_2^2\boldsymbol{e}_1.
\end{equation}
We assume that fluid inertia is neglected, i.e.~that the local particle Reynolds number $\Rey_\text{p,L}(x_\text{CM,2})=\rho_\text{f}\dot{\gamma}'l^3|x_\text{CM,2}|/\mu$ is zero ($\rho_\text{f}$ is the fluid density, $\mu$ is the fluid dynamic viscosity), so that the flow is governed by the incompressible Stokes equations
\begin{equation}
\label{eq:StEq}
\left. \begin{array}{c}
  \bnabla \bcdot \boldsymbol{u}=0,\\[5pt]
  \bnabla p=\nabla^2\boldsymbol{u},
\end{array}\right\}
\end{equation}
which are non-dimensionalized using characteristic length $l$, time $(\dot{\gamma}'l)^{-1}$ and pressure $\mu \dot{\gamma}'l$. The boundary conditions of the problem is that the background flow is obtained far away from the particle, i.e.
\begin{equation}
  \lim_{\lvert\boldsymbol{x}\rvert \to \infty} \Big(\boldsymbol{u}(\boldsymbol{x},t)-\boldsymbol{u}_\text{bg}(\boldsymbol{x})\Big)=\boldsymbol{0},
\end{equation}
and that there is no slip of the fluid on the particle surface $\Gamma$, i.e.
\begin{equation}
\boldsymbol{u}(\boldsymbol{x},t)=\boldsymbol{V}+\boldsymbol{\omega}\times(\boldsymbol{x}-\boldsymbol{x}_\text{CM}), \qquad \boldsymbol{x}\in\Gamma,
\end{equation}
where $\boldsymbol{V}$ is the (non-dimensional) velocity of the particle centre-of-mass located at $\boldsymbol{x}_\text{CM}$ and the (non-dimensional) particle angular velocity is $\boldsymbol{\omega}$. The motion of the particle is determined by the equations of motion
\begin{eqnarray}
\label{eq:GovEqMotion}
\left. \begin{array}{c}
\boldsymbol{F}=\textit{St}_\text{trans.}\Phi\dot{\boldsymbol{V}},\\[5pt]
\boldsymbol{M}=\textit{St}_\text{rot.}\left[\mathsfbi{I}\dot{\boldsymbol{\omega}}+\boldsymbol{\omega}\times(\mathsfbi{I}\boldsymbol{\omega})\right],
\end{array}\right\}
\end{eqnarray}
where $\textit{St}_\text{trans.}$ and $\textit{St}_\text{rot.}$ are the Stokes numbers characterizing the translational inertia and rotational inertia, respectively. We initially set these to the same value, corresponding to a spheroid with uniform density $\rho_\text{p}$. Both translational and rotational inertia will thus be set using the global Stokes number $\textit{St}_\text{G}$:
\begin{equation}
\textit{St}_\text{G}=\textit{St}_\text{trans.}=\textit{St}_\text{rot.}=\frac{\rho_\text{p}}{\rho_\text{f}}\Rey_\text{p,L}(x_\text{CM,2}=1)=\frac{\rho_\text{p}\dot{\gamma}'l^3}{\mu}.
\end{equation}
The parameter $\Phi=(4/3)\upi r_\text{p}^{-2}$ is the non-dimensional volume of the spheroidal particle. The non-dimensional force, torque and inertial tensor are given by $\boldsymbol{F}$, $\boldsymbol{M}$ and $\mathsfbi{I}$, respectively. The equations are non-dimensionalized with characteristic torque $\mu\dot{\gamma}'l^4$ and the characteristic inertial tensor element $\rho_\text{p} l^5$. Note that in order for the local $\Rey_\text{p,L}(x_\text{CM,2})$ to be negligible, while the local $\textit{St}_\text{L}(x_\text{CM,2})=\rho_\text{p}\dot{\gamma}'l^3|x_\text{CM,2}|/\mu$ is relevant, the solid-to-fluid density ratio must fulfill $\alpha=\rho_\text{p}/\rho_\text{f}\gg1$. Of course, this condition causes the particle to sediment in the presence of gravity. Initially in this study, we will neglect gravity effects but we will discuss these effects in Section~\ref{sec:GravForceEffects}.

\section{Method}\label{sec:Method}

\subsection{Boundary integral formulation}

For $\textit{St}_\text{G} > 0$, the equations of motion \eqref{eq:GovEqMotion} are solved using Matlab's ordinary differential equation solver \texttt{ode113} with relative tolerance $10^{-8}$. The force~$\boldsymbol{F}$ and torque~$\boldsymbol{M}$ are computed numerically from the position, orientation and velocity of the particle at every time step using a boundary integral formulation based on \citet{PowerMiranda87} and \citet{Gonzalez09}. In this formulation, the flow field in the fluid domain~$D_\text{f}$ is expressed as integrals over the particle surface~$\Gamma$:
\begin{equation}
  \label{eq:flow}
  \boldsymbol{u}(\boldsymbol{x},t) = \boldsymbol{\mathcal{D}}[\Gamma,\boldsymbol{q}](\boldsymbol{x}) + \boldsymbol{\mathcal{V}}[\boldsymbol{x}_\text{CM},\boldsymbol{F},\boldsymbol{M}]({\boldsymbol{x}}) +
   \boldsymbol{u}_\text{bg}(\boldsymbol{x}), \qquad  \boldsymbol{x} \in D_\text{f}.
\end{equation}
Here, the Stokes double layer potential $\boldsymbol{\mathcal{D}}$ with density $\boldsymbol{q}$ is given by (Einstein's summation convention is used here, with all indices ranging over $\{1,2,3\}$)
\begin{equation}
  \label{eq:dbl}
  \mathcal{D}_{i}[\Gamma,\boldsymbol{q}](\boldsymbol{x}) = \int_\Gamma T_{ijk}(\boldsymbol{x}-\boldsymbol{y})  q_j(\boldsymbol{y}) n_k(\boldsymbol{y}) \, \mathrm{d}S_{\boldsymbol{y}}, \qquad
  T_{ijk}(\boldsymbol{r}) = -6  \frac{r_ir_jr_k}{\lvert\boldsymbol{r}\rvert^5}.
\end{equation}
The completion flow $\boldsymbol{\mathcal{V}}$ is needed to represent the force and torque, and is given by
\begin{gather}
  \mathcal{V}_{i}[\boldsymbol{x}_\text{CM},\boldsymbol{F},\boldsymbol{M}](\boldsymbol{x}) = \frac{1}{8\upi} \left(C_{ij}(\boldsymbol{x}-\boldsymbol{x}_\text{CM}) F_{j} + H_{ij}(\boldsymbol{x},\boldsymbol{x}_\text{CM}) M_{j}\right), \\
  C_{ij}(\boldsymbol{r}) = \frac{\delta_{ij}}{\lvert\boldsymbol{r}\rvert} + \frac{r_{i}r_{j}}{\lvert\boldsymbol{r}\rvert^3}, \qquad H_{ij}(\boldsymbol{x},\boldsymbol{y}) = \frac{\varepsilon_{ijk}r_{k}}{\lvert\boldsymbol{r}\rvert^3},\nonumber
\end{gather}
where $\delta_{ij}$ is the Kronecker delta and $\varepsilon_{ijk}$ is the alternating symbol.

By letting $\boldsymbol{x}$ go to $\Gamma$ in \eqref{eq:flow} and using the no-slip condition together with a jump property of $\boldsymbol{\mathcal{D}}$, one arrives at a boundary integral equation for $\boldsymbol{q}$:
\begin{equation}
  \label{eq:bie}
  -4\upi\boldsymbol{q}(\boldsymbol{x}) + \boldsymbol{\mathcal{D}}[\Gamma,\boldsymbol{q}](\boldsymbol{x}) + \boldsymbol{\mathcal{V}}[\boldsymbol{x}_\text{CM},\boldsymbol{F},\boldsymbol{M}](\boldsymbol{x})
  \boldsymbol{V} + \boldsymbol{\omega} \times (\boldsymbol{x} - \boldsymbol{x}_\text{CM}) - \boldsymbol{u}_\text{bg}(\boldsymbol{x}),
  \quad \boldsymbol{x} \in \Gamma,
\end{equation}
which is closed using the relations
\begin{equation}
  \label{eq:forcrel}
  \boldsymbol{F} = \int_\Gamma \boldsymbol{q}(\boldsymbol{y}) \, \mathrm{d}S_{\boldsymbol{y}}, \qquad
  \boldsymbol{M} = \int_\Gamma \boldsymbol{q}(\boldsymbol{y}) \times (\boldsymbol{y} - \boldsymbol{x}_\text{CM}) \, \mathrm{d}S_{\boldsymbol{y}}.
\end{equation}
For details we refer to \citet{Klinteberg16}.

For $\textit{St}_\text{G}=0$, which corresponds to a massless particle, the solution procedure is different since we know from \eqref{eq:GovEqMotion} that $\boldsymbol{F} = \boldsymbol{M} = \boldsymbol{0}$. In this case, \eqref{eq:bie} must instead be solved for the velocities $\boldsymbol{V}$ and $\boldsymbol{\omega}$, for which relations similar to \eqref{eq:forcrel} hold (see \citet{Klinteberg16}). 

Hence, the flow field produced by \eqref{eq:flow} is the solution to the Stokes equations \eqref{eq:StEq}, given that the density $\boldsymbol{q}$ solves \eqref{eq:bie}.

\subsection{Discretization and quadrature by expansion}\label{sec:DiscQuad}

\begin{figure}
  \centering
  \includegraphics[trim=8.9cm 14cm 8.7cm 13.75cm,clip]{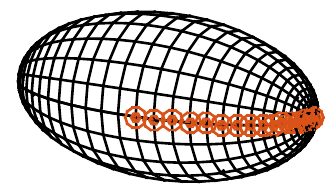}
  \caption{The spheroidal grid. Due to symmetry it is enough to store precomputed matrices for the $n_{\theta}/2$ grid points along the first half meridian, marked with circles.}
  \label{fig:qbx-precomp}
\end{figure}

The boundary integral equation \eqref{eq:bie} is discretized using the Nystr\"{o}m method \citep[ch.~4]{atkinson97}, which enforces the equation at the grid points of the discretized particle
surface, shown in figure~\ref{fig:qbx-precomp}. We use the trapezoidal rule with equidistant points in the periodic direction (along a circle of latitude) and an $n_{\theta}$-point Gauss--Legendre quadrature rule in the non-periodic direction (along a meridian). This choice gives us spectral accuracy for smooth and well-resolved integrands on the particle surface \citep{Klinteberg16}.

Applying the quadrature rule directly to the integral equation \eqref{eq:bie} is problematic for two reasons. Firstly, the integration kernel of the double layer potential \eqref{eq:dbl}
is singular when the evaluation point $\boldsymbol{x}$ coincides with a point $\boldsymbol{y}$ on the boundary, and can therefore not be handled by a quadrature rule for smooth functions. Moreover, when \eqref{eq:flow} is evaluated in a point $\boldsymbol{x}$ which is close to the boundary, but not on it, the integration kernel of \eqref{eq:dbl} is sharply peaked, which causes a significant loss of accuracy close to the particle surface. We use a recent method called quadrature by expansion (QBX) to treat both of these problems. The idea behind this method is described briefly below; for a detailed description, we refer to \citet{Klinteberg16}.

\begin{figure}
  \centering
  \includegraphics{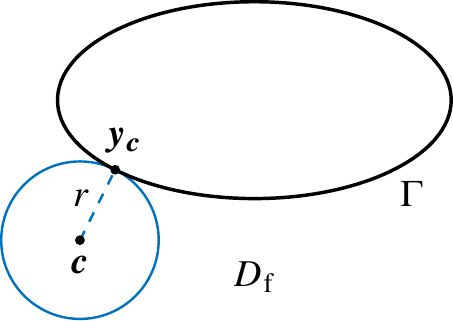}\hspace{4em}%
  \includegraphics{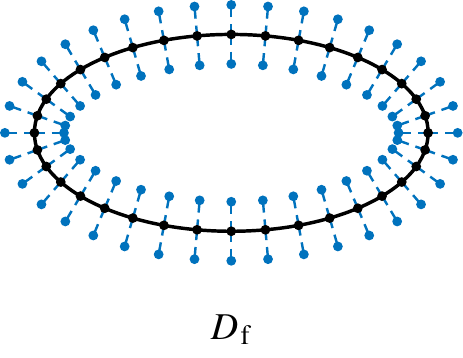}
  \caption{Left: An expansion around a point $\boldsymbol{c}$ is valid within its ball of convergence of radius $r$ and can be used to evaluate the double layer potential at exactly one point
  $\boldsymbol{y_c}$ on the boundary. Right: For every grid point on the boundary we create an inner and an outer expansion centre.}
  \label{fig:qbx}
\end{figure}

QBX is based on the observation that the double layer potential $\boldsymbol{\mathcal{D}}$ is a smooth function away from the boundary. To avoid the problems close to the boundary, we can create a local expansion of $\boldsymbol{\mathcal{D}}$ around a point $\boldsymbol{c}$ away from the boundary. This expansion can be used to evaluate $\boldsymbol{\mathcal{D}}$ at exactly one point on the boundary, as shown in figure~\ref{fig:qbx}. To solve the integral equation we thus need one expansion centre for every grid point. Since $\boldsymbol{\mathcal{D}}$ has different limits from the interior and the exterior, we use both an inner and an outer expansion centre and compute the potential
$\boldsymbol{\mathcal{D}}^\text{QBX}$ as an average to get the value on $\Gamma$.

There exist matrices $\mathsfbi{D}_j$ such that $\boldsymbol{\mathcal{D}}^\text{QBX}[\Gamma,\boldsymbol{q}](\boldsymbol{x}_j) = \mathsfbi{D}_j \boldsymbol{Q}$ for each grid point $\boldsymbol{x}_j$ on the surface, where $\boldsymbol{Q}$ is a vector containing the values of the density~$\boldsymbol{q}$ in all grid points. The matrices $\mathsfbi{D}_j$ depend only on the geometry of the spheroid and can be precomputed; the result can be seen as a regular but target-specific quadrature rule. Due to rotational and mirror symmetry, it is sufficient to store matrices for the $n_{\theta}/2$ grid points along the first half meridian, shown in fig.~\ref{fig:qbx-precomp}. The precomputation allows the method to be both fast and accurate, since precomputation can be done once with high accuracy and the result is reused in every time step.

\subsection{Validation}

The method has been validated against test cases for both inertial and massless particles, in linear shear flow and the quadratic background flow considered here. The analytical solutions in these cases are provided by \citet{Jeffery} and \citet{Chwang}; the validation is further described in \citet{bagge16}. Using the parameters given in appendix~\ref{appA}, the relative errors are below $10^{-6}$ in all test cases.

\section{Results}\label{sec:Results}

\begin{figure}
\begin{center}
\includegraphics{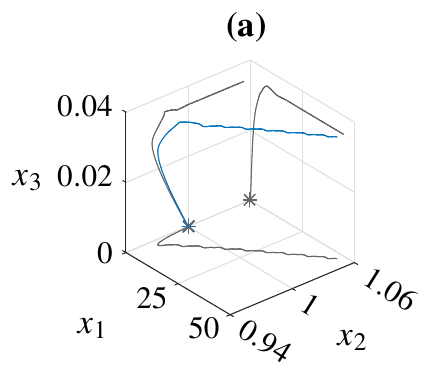}\hfill%
\includegraphics{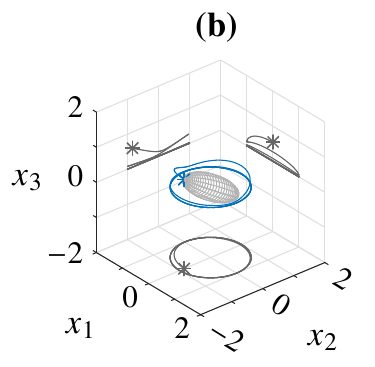}\hfill%
\includegraphics{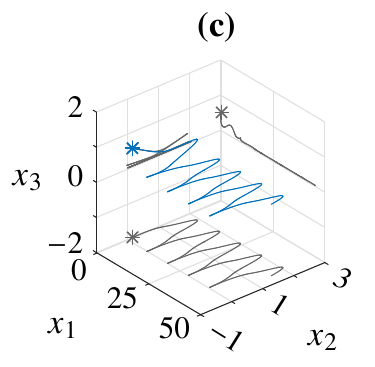}
\caption{\label{fig:DriftExample}Trajectories of a particle with $r_\text{p}=3$ and $\textit{St}_\text{G}=50$, initialized at rest with an oblique orientation at $\boldsymbol{x}_\text{CM} = (0,1,0)$; (a) trajectory of centre
  $\boldsymbol{x}_\text{CM}$; (b) trajectory of endpoint $\boldsymbol{s}$, ignoring translation of the centre; (c) trajectory of endpoint including translation, i.e~$\boldsymbol{x}_\text{CM}+\boldsymbol{s}$.}
\end{center}
\end{figure}

We start by considering a prolate spheroid of $r_\text{p}=3$ and $\textit{St}_\text{G}=50$ initialized at rest at position $\boldsymbol{x}_\text{CM}=(0,1,0)$ with a slightly oblique orientation. The trajectories of the particle translation and orientation are illustrated in figure~\ref{fig:DriftExample}. The orientation of the particle is clearly drifting towards a rotation around its minor axis, an intermittent rotation that we call \emph{tumbling}. This is the same type of behaviour seen in a simple shear flow by \citet{LundellCarlsson}. What is more striking is what is happening to the translational motion of the particle. After an initial transient the particle ends up \emph{laterally drifting towards regions of higher shear.}

\begin{figure}
\begin{center}
  \includegraphics{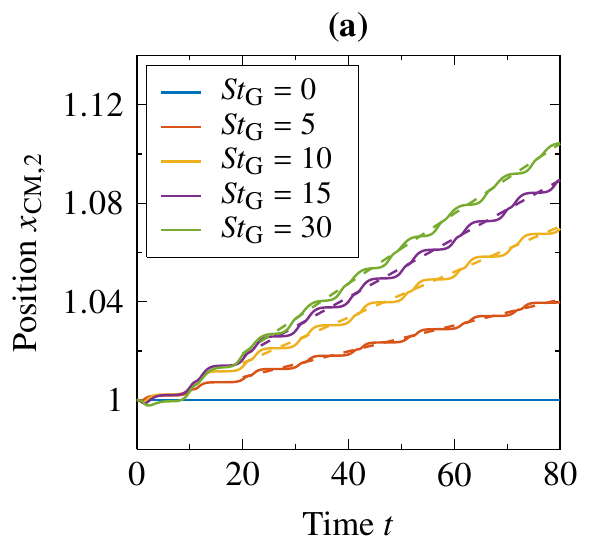}\hfill%
  \includegraphics{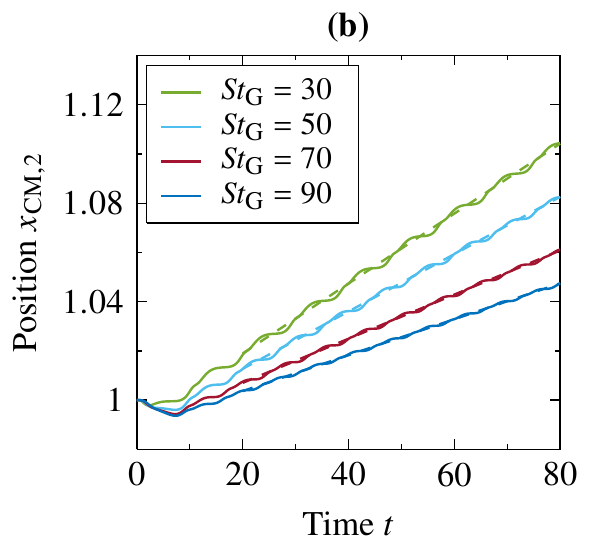}
  \caption{\label{fig:Drift}Illustration of the inertial drift of a particle with $r_\text{p}=3$ at (a) $\textit{St}_\text{G} \leq 30$ and (b) $\textit{St}_\text{G} \geq 30$.}
\end{center}
\end{figure}

To quantify the observed drift, we consider a particle of $r_\text{p}=3$ that is initialized at position $\boldsymbol{x}_\text{CM}=(0,1,0)$ with velocity $\boldsymbol{V}=\boldsymbol{u}_\text{bg}(\boldsymbol{x}_{\text{CM}})$, aligned in the flow gradient direction, i.e.~at $\boldsymbol{s}=(0,1,0)$ with zero angular velocity. Note that at the given initial position, the local and global Stokes numbers are the same, i.e.~$\textit{St}_\text{G}=\textit{St}_\text{L}(x_\text{CM,2}=1)$. The particle is then free to both rotate and translate up until $t=80$. This is repeated for a number of different $\textit{St}_\text{G}$ in the range $\textit{St}_\text{G}\in[0,100]$. The resulting trajectory of the particle centre-of-mass is shown in figure~\ref{fig:Drift}. While the spheroid with no inertia ($\textit{St}_\text{G}=0$) does not show any lateral motion at all, as soon as there is particle inertia, the particle starts to migrate in positive $x_2$-direction, i.e.~towards higher shear. Note, that the trajectories are plotted as function of the global time scale even though the local relevant time scale is changing (due to increasing local shear) with $x_\text{CM,2}$. However, this effect is negligible since the particle has not moved significantly in the $x_2$-direction at $t=80$. For all $\textit{St}_\text{G}$ studied here, the particle assumes its final state (a periodic rotation) quite quickly after an initial transient while the particle is accelerating to the surrounding flow. The final average lateral drift velocity $V_\text{drift}=\langle V_2\rangle$ for each $\textit{St}_\text{G}$ is seen to be constant, and is estimated by fitting a linear function to $x_{\text{CM},2}(t)$ between times $t=20$ and $80$ as seen by the dashed lines in fig.~\ref{fig:Drift}. From the figure it is also evident that there is a critical Stokes number $\textit{St}_\text{c}\approx 30$ such that $V_\text{drift}$ has a maximum value when $\textit{St}_\text{G}=\textit{St}_\text{c}$. At $\textit{St}_\text{G}\leq30$ (fig.~\ref{fig:Drift}a) the drift velocity $V_\text{drift}$ is increasing with $\textit{St}_\text{G}$ while at $\textit{St}_\text{G}\geq 30$ (fig.~\ref{fig:Drift}b), $V_\text{drift}$ is decreasing with $\textit{St}_\text{G}$.

\begin{figure}
\begin{center}
  \includegraphics{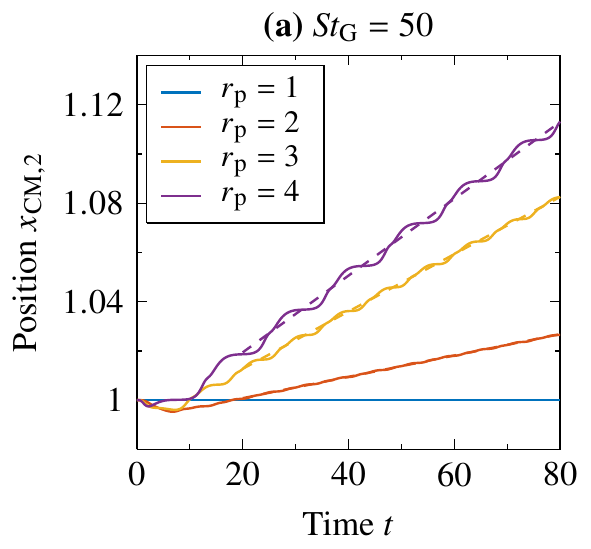}\hfill%
  \includegraphics{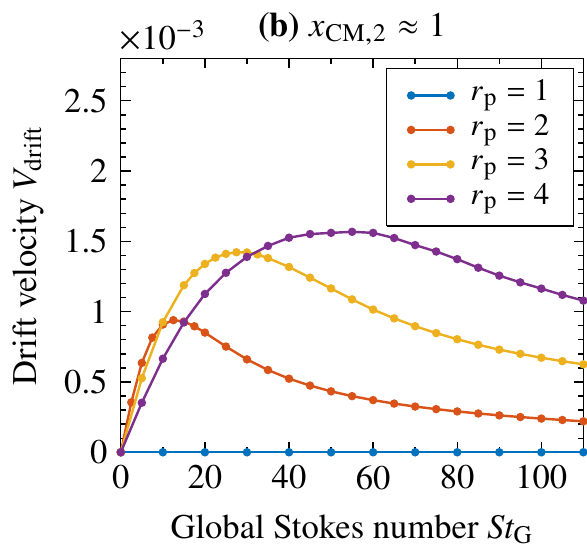}
  \caption{\label{fig:rpDep}Inertial drift velocity $V_\text{drift}$ as a function of aspect ratio $r_\text{p}$; (a) position versus time; (b) $V_\text{drift}$ versus $\textit{St}_\text{G}$ at various $r_\text{p}$.}
\end{center}
\end{figure}

Considering now a particle with different aspect ratios $r_\text{p}=1,2,3,4$ at $\textit{St}_\text{G}=50$ in figure~\ref{fig:rpDep}a, we find that a spherical particle, as expected, has no lateral drift. As soon as the particle gets a prolate shape, it gets a constant drift velocity, which at $\textit{St}_\text{G}=50$ is increasing with $r_\text{p}$. In fig.~\ref{fig:rpDep}b, the final drift velocity $V_\text{drift}$ is plotted as function of $\textit{St}_\text{G}$ for the different $r_\text{p}$ and we find that both the maximum $V_\text{drift,max}$ and the Stokes number $\textit{St}_\text{c}$ where this occurs are indeed increasing with $r_\text{p}$.

\begin{figure}
\begin{center}
  \includegraphics{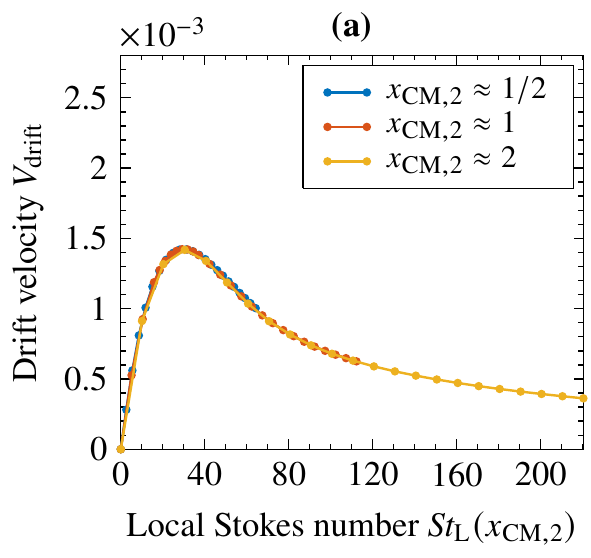}\hfill%
  \includegraphics{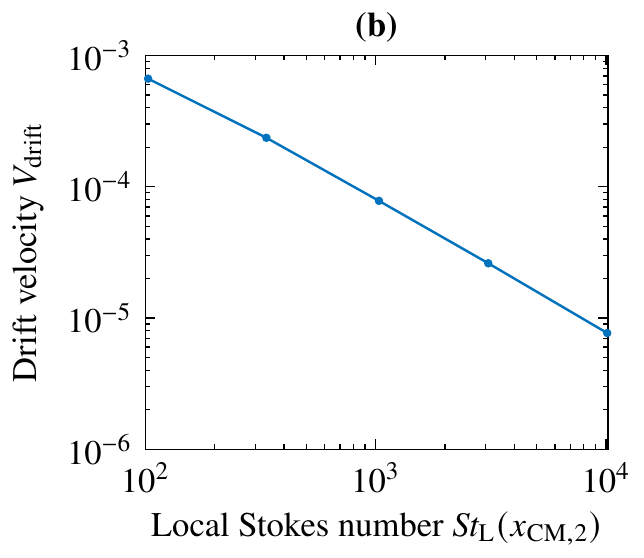}
  \caption{\label{fig:posDep}(a) The drift velocity $V_\text{drift}$ as a function of  local $\textit{St}_\text{L}(x_\text{CM,2})$ for different initial heights and $r_p=3$; figure shows that $V_\text{drift}$ is only dependent on the local $\textit{St}_\text{L}(x_\text{CM,2})$, with a maximum at $\textit{St}_\text{L}(x_\text{CM,2,c}) = \textit{St}_\text{c}$; (b) the drift velocity $V_\text{drift}$ at very high values of $\textit{St}_\text{L}$.}
\end{center}
\end{figure}

As previously mentioned, if we would release the particle at different heights $x_\text{CM,2}$ in the flow, the particle would experience different shear rates and the relevant time scale would change. In this case the local $\textit{St}_\text{L}=|x_\text{CM,2}|\textit{St}_\text{G}$ depends on lateral position in the flow, which in turn determines the drift velocity. To demonstrate this, we released the particle of $r_\text{p}=3$ at two more locations $x_\text{CM,2}=1/2$ and $x_\text{CM,2}=2$. We can see in figure~\ref{fig:posDep}a that $V_\text{drift}$ is only dependent on $\textit{St}_\text{L}$ as all the curves collapse when the velocity is plotted versus this parameter. Consequently, there will always be a particle position in the channel $x_\text{CM,2}=x_\text{CM,2,c}$ where we have a maximum drift velocity, which is where $\textit{St}_\text{L}(x_\text{CM,2,c})=\textit{St}_\text{c}$. If we e.g.~start at $|x_\text{CM,2}|<1$, we will always eventually reach this position since the particle is drifting towards regions of higher shear. Eventually, the drift will vanish as $\textit{St}_\text{L}\rightarrow \infty$, which is demonstrated in fig.~\ref{fig:posDep}b where the results of $V_\text{drift}$ at very high $\textit{St}_\text{L}$ are illustrated.

\section{Discussion}\label{sec:Discussion}

The previous section has shown the results that particle inertia is sufficient to cause a lateral drift of a particle in a quadratic velocity profile. The source of the resulting translational and rotational motion will be discussed in this section.

\subsection{Without inertia}
\citet{Jeffery} and \citet{lamb1932hydrodynamics} present expressions for the force $\boldsymbol{F}_0$ and torque $\boldsymbol{M}_0$ on a translating and rotating spheroidal particle in a quiescent fluid, while \citet{Chwang} derived expressions for the force $\boldsymbol{F}^\text{Chw}$ and torque $\boldsymbol{M}^\text{Chw}$ on a non-translating and non-rotating spheroidal particle in a parabolic velocity profile. One important thing to note is that the results by \citet{Chwang} were derived for a paraboloidal profile $\boldsymbol{u}^\text{Chw}_\text{bg}(\boldsymbol{x})=(1/2)(x_2^2+x_3^2)\boldsymbol{e}_1$. By adjusting these results, we can express the force $\boldsymbol{F}^\text{par}$ and torque $\boldsymbol{M}^\text{par}$ valid for the here considered parabolic background velocity profile $\boldsymbol{u}_\text{bg}(\boldsymbol{x})=(1/2)x_2^2\boldsymbol{e}_1$. This derivation is given in appendix~\ref{appB}.

Through the principle of superposition, the total force and torque on a moving and rotating spheroidal particle in a parabolic velocity profile in the absence of fluid inertia can be found by combining the adjusted results by \citet{Chwang} with the expressions of \citet{Jeffery} and \citet{lamb1932hydrodynamics}, i.e.~$\boldsymbol{F}=\boldsymbol{F}_0+\boldsymbol{F}^\text{par}$ and $\boldsymbol{M}=\boldsymbol{M}_0+\boldsymbol{M}^\text{par}$. If we want to study the free motion of the spheroid with no particle inertia ($\textit{St}_\text{trans.}=\textit{St}_\text{rot.}=0$ in \eqref{eq:GovEqMotion}), the resulting force $\boldsymbol{F}$ and torque $\boldsymbol{M}$ are set to zero.

\subsubsection{Rotation without inertia}

Let us now consider a particle that is oriented in the flow-gradient plane ($\theta=\upi/2$). The torque on a spheroidal particle rotating in the flow-gradient plane with angular velocity $\dot{\phi}$ in a quiescent fluid is given in non-dimensional form by \citet{Jeffery} as
\begin{equation}
\label{eq:LambTorque}
\boldsymbol{M}_0(\dot{\phi}) = -\frac{16\upi}{3} \frac{r_\text{p}^2 + 1}{K_1 r_\text{p}^2 + K_2} \dot{\phi} \, \boldsymbol{e}'_2,
\end{equation}
where
\begin{equation}
  K_1 = \int_0^\infty
  \frac{\mathrm{d}\beta}{(1+\beta)^{3/2}(r_\text{p}^{-2} + \beta)},
  \qquad
  K_2 = \int_0^\infty
  \frac{\mathrm{d}\beta}{(1+\beta)^{1/2}(r_\text{p}^{-2} + \beta)^2}.
\end{equation}
The torque on a stationary particle with orientation $\phi$ and location $\boldsymbol{x}_\text{CM}$ in a parabolic velocity profile is given through the (non-dimensional) adjusted expression by \citet{Chwang} (see appendix~\ref{appB}) as
\begin{equation}
\label{eq:ChwangTorque}
\boldsymbol{M}^\text{par}(\phi,\boldsymbol{x}_\text{CM})=-\frac{32\upi}{3} \frac{E^3 (1 - E^2 \cos^2 \phi)}{-2E+(1+E^2) \log[(1+E)/(1-E)]} x_{\text{CM},2} \boldsymbol{e}'_2,
\end{equation}
where $E = \sqrt{1-r_\mathrm{p}^{-2}}$ is the eccentricity of the spheroid.

The total torque on the particle rotating arbitrarily in a parabolic velocity profile is given by $\boldsymbol{M}=\boldsymbol{M}_0+\boldsymbol{M}^\text{par}$. The resulting expression is actually exactly equivalent to the expression by \citet{Jeffery} in a linear shear flow. The conclusion drawn already by \citet{Chwang} is thus that the quadratic terms in the background flow do not affect the particle rotation. Consequently, the particle without rotational inertia ($\textit{St}_\text{rot.}=0$), which rotates according to the solution of $\boldsymbol{M}=\boldsymbol{0}$, will just rotate according to the local shear rate in an intermittent tumbling motion described by \citet{Jeffery} as
\begin{equation}
\label{eq:JefferyAngVel}
\dot{\phi}(\phi,\boldsymbol{x}_\text{CM}) = |x_{\text{CM},2}| \frac{r_\text{p}^2 \sin^2 \phi + \cos^2 \phi}{r_\text{p}^2+1}.
\end{equation}
The rotational period is given by
\begin{equation}
\label{eq:JefferyPeriod}
 T_J=\frac{2\upi}{|x_\text{CM,2}|}(r_p+r_p^{-1}).
\end{equation}

\subsubsection{Translation}\label{sec:TranslationDiscussion}

The force on a spheroidal particle moving with instantaneous velocity $\boldsymbol{V}$ and orientation $\phi$ in a quiescent fluid is given by \citet{lamb1932hydrodynamics} as
\begin{equation}
\label{eq:LambForce}
\boldsymbol{F}_0(\boldsymbol{V},\phi)=-16\upi \mathsfbi{K}(\phi)\boldsymbol{V},
\end{equation}
where the tensor $\mathsfbi{K}$ in the body-fixed coordinate system (equivalent to $\phi=0$) is given by
\begin{equation}
  \mathsfbi{K}(0) = \left(\begin{matrix}
    (K_0+K_1)^{-1} & 0 & 0 \\
    0 & (K_0+K_2 r_\text{p}^{-2})^{-1} & 0 \\
    0 & 0 & (K_0+K_2 r_\text{p}^{-2})^{-1} \\
  \end{matrix}
\right)
\end{equation}
with
\begin{equation}
  K_0 = \int_0^\infty
  \frac{\mathrm{d}\beta}{(1+\beta)^{1/2}(r_\text{p}^{-2} + \beta)}.
\end{equation}
If $\phi\neq 0$, the tensor will simply be transformed according to $\mathsfbi{K}(\phi)=\mathsfbi{R}(\phi)\mathsfbi{K}(0)\mathsfbi{R}^{-1}(\phi)$ with the rotation matrix $\mathsfbi{R}$.

The force on a stationary particle with orientation $\phi$ (between $\boldsymbol{s}$ and $\boldsymbol{u}_\text{CM}$) and location $\boldsymbol{x}_\text{CM}$ in a parabolic velocity profile is given through the adjusted expression by \citet{Chwang} (see appendix~\ref{appB}) as
\begin{eqnarray}
\label{eq:ChwangForce}
\boldsymbol{F}^\text{par}(\phi,\boldsymbol{x}_\text{CM}) &=& \frac{8 \upi E^3 \cos \phi}{3}
  \frac{3 x_{\text{CM},2}^2 + 1 - E^2 \cos^2 \phi}{-2E + (1+E^2) \log[(1+E)/(1-E)]} \boldsymbol{e}'_1 + \\
  &&- \frac{16\upi E^3 \sin \phi}{3} \frac{3 x_{\text{CM},2}^2 + 1 - E^2 \cos^2 \phi}{2E + (3E^2-1) \log[(1+E)/(1-E)]} \boldsymbol{e}'_2.\nonumber
\end{eqnarray}
The total force on the particle arbitrarily translating and rotating in a parabolic velocity profile is given by $\boldsymbol{F}=\boldsymbol{F}_0+\boldsymbol{F}^\text{par}$. If the particle is  oriented with an angle $\phi$ and free to translate in the absence of translational inertia ($\textit{St}_\text{trans.}=0$), the velocity that satisfies $\boldsymbol{F}=\boldsymbol{0}$ is given by
\begin{equation}
\label{eq:VelocityByChwang}
\boldsymbol{V}(\phi,\boldsymbol{x}_\text{CM}) = \frac{1}{6} \left( 3 x_{\text{CM},2}^2 + 1 - E^2 \cos^2 \phi \right) \boldsymbol{e}_1.
\end{equation}
The particle has a velocity in the flow direction, which depends on the instantaneous angle $\phi$. If the particle is free to rotate without rotational inertia, it will perform an intermittent tumbling motion according to \eqref{eq:JefferyAngVel}. Consequently, this thus results in an intermittent translational motion in the flow direction.

\subsection{Adding inertia}

\subsubsection{Only rotational inertia}

\begin{figure}
\begin{center}
\includegraphics{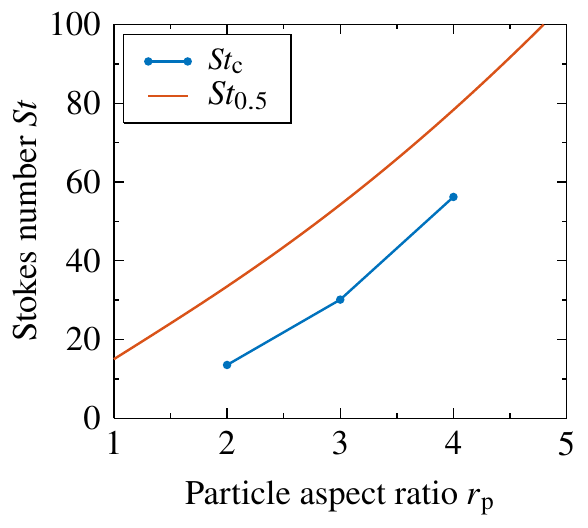}
  \caption{\label{fig:CompareSimpleShear}The critical Stokes numbers $\textit{St}_{c}$ and $\textit{St}_{0.5}$ as functions of $r_\text{p}$.}
\end{center}
\end{figure}

Although not easily achievable in practice, let us now consider the case where $\textit{St}_\text{rot.}>0$ but translational inertia is neglected $\textit{St}_\text{trans.}=0$. Since we know from the expressions above that the translational motion is only in the flow-direction in the absence of translational inertia and the rotational motion is only dependent on the motion in the gradient-direction, the translation will not influence the particle rotation. With added rotational inertia, the particle will thus behave exactly according to \citet{LundellCarlsson}, where the instantaneous torque during the final rotational motion typically is non-zero but fulfilling $\int_0^T \boldsymbol{M} \, \mathrm{d}t=\boldsymbol{0}$ for the rotational period $T$. At extremely large $\textit{St}_\text{rot.}\rightarrow \infty$, the particle will rotate with a constant angular velocity $\dot{\phi}=-0.5|x_\text{CM,2}|$ and period $T_H=4\upi/(|x_\text{CM,2}|)$ as the particle inertial forces overcome viscous forces from the surrounding fluid. A critical rotational Stokes number $\textit{St}_{0.5}$ was introduced by \citet{LundellCarlsson} to describe the transition. This number is defined as the $\textit{St}_\text{rot.}$ where the tumbling period is  $T=(T_J+T_H)/2$. With corrections given by \citet{Nilsen}, this number can be found through
\begin{equation}
\label{eq:St05}
\textit{St}_{0.5}=\frac{20[0.7+0.3[1-\varepsilon^2]^{5/8})]}{\Lambda[K_1+\Lambda K_2]},
\end{equation}
with
\[
  \varepsilon=\frac{r_\text{p}^{2}-1}{r_\text{p}^{2}+1}, \qquad
  \Lambda=\frac{1-\varepsilon}{1+\varepsilon}.
\]
\citet{LundellCarlsson} also found that the particle initially oriented out of the flow-gradient plane always still drifted towards the tumbling motion due to the centrifugal forces on the particle. The rate of this orbit drift was seen to be close to maximum when $\textit{St}_\text{rot.}=\textit{St}_{0.5}$. Interestingly, the critical rotational Stokes number $\textit{St}_{0.5}$ for maximum orbit drift seems to scale similarly with aspect ratio as the critical translational Stokes number $\textit{St}_\text{c}$ for the lateral drift as illustrated in figure~\ref{fig:CompareSimpleShear}. This indicates that the maximum lateral drift also arises in the transition when inertial forces overcome viscous damping.

\begin{figure}
\begin{center}
\includegraphics{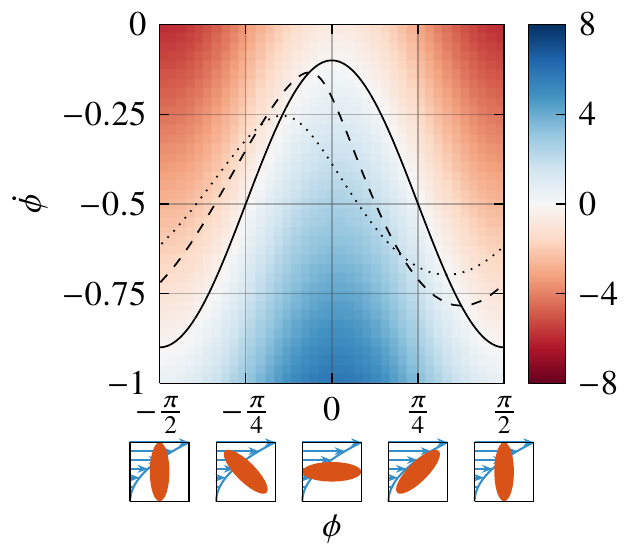}
\caption{\label{fig:CompareTorque}Torque $M_z$ given on a particle ($r_\text{p}=3$) in a quadratic flow at $x_\text{CM,2}=1$ with no translational velocity in a tumbling orbit ($s_3=\cos\theta=0$) as function of orientation $\phi$ and angular velocity $\dot{\phi}$; the solid, dashed and dotted lines shows the superimposed path of a particle that is free to translate at $\textit{St}_\text{L} = 0$, $50$ and $100$, respectively.}
\end{center}
\end{figure}

The fact that the instantaneous torque on the particle is exactly represented by the solutions by \citet{Jeffery} is also confirmed by the Stokes flow simulations in the present work. In figure~\ref{fig:CompareTorque}, we plot the torque $M_3$ on the particle as a function of orientation $\phi$  and angular velocity $\dot{\phi}$. Furthermore, we superimpose the trajectories of the numerical simulations from our results at higher $\textit{St}_\text{L}=\textit{St}_\text{rot.}=\textit{St}_\text{trans.}$. Here we can clearly see how the particle travels through regions of both positive and negative torque during a rotation and approaching constant angular velocity $\dot{\phi}=-0.5|x_\text{CM,2}|$ as $\textit{St}_\text{L}$ increases. We find in the numerical simulations also that the translational inertia has no effect on the rotation, except for the fact that the lateral drift changes the local time scale (through the local shear rate) and thus also the local Stokes number $\textit{St}_\text{L}$.

\subsubsection{Adding translational inertia}
With no translational inertia, the particle moves with oscillating $V_1$-velocity ($V_2=0$) corresponding to $\boldsymbol{F}=\boldsymbol{0}$. The oscillation period corresponds to the oscillation period of $\phi$.

With added rotational inertia (still no translational inertia) as we observed previously, there will be a different oscillation period, but $V_1$ is still given by the $\boldsymbol{F}=\boldsymbol{0}$ solution, which does not imply any lateral drift ($V_2=0$). With added translational inertia (no rotational inertia) the particle will be slow to react to the forces and the particle will typically experience non-zero forces but eventually leading to a translational motion that fulfills $\int_0^T \boldsymbol{F} \, \mathrm{d}t=\boldsymbol{0}$. This solution actually has $V_2\neq 0$ and the particle will drift laterally.

\begin{figure}
\begin{center}
  \includegraphics{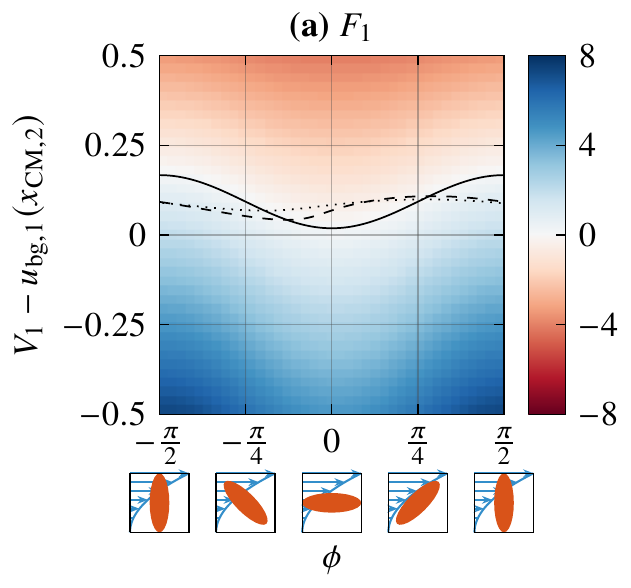}\hfill%
  \includegraphics{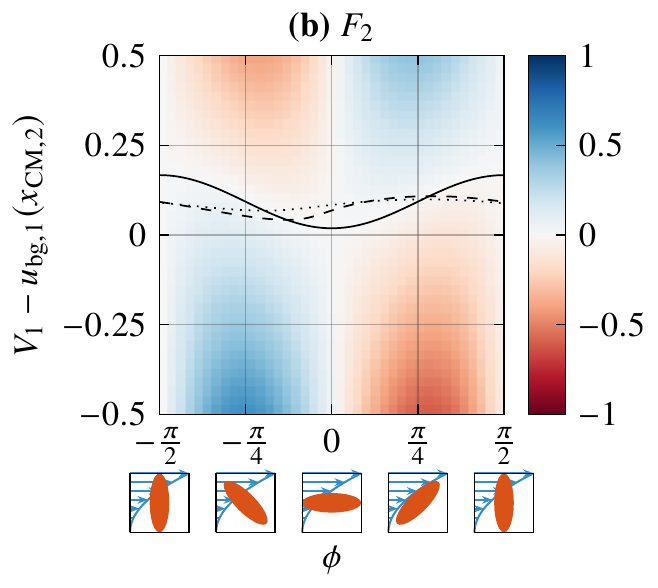}
\caption{\label{fig:CompareForce}Forces given on a particle ($r_p=3$) with fixed orientations of a tumbling orbit ($s_3=\cos\theta=0$) as functions of orientation $\phi$ and streamwise velocity $V_1$; the solid, dashed and dotted lines show the  superimposed path of a particle that is free to translate at $\textit{St}_\text{L} = 0$, $50$ and $100$, respectively; (a) streamwise force component $F_1$; (b) lateral force component $F_2$.}
\end{center}
\end{figure}

With added rotational inertia, the oscillation period of the velocity will decrease as the oscillation period of $\phi$ will decrease. The fact that the instantaneous force on the particle is exactly represented by the analytical expressions of $\boldsymbol{F}=\boldsymbol{F}_0+\boldsymbol{F}^\text{par}$ is confirmed in the present Stokes flow simulations. In figure~\ref{fig:CompareForce}, we illustrate the instantaneous streamwise and lateral forces $F_1$ and $F_2$ for a given orientation $\phi$ and streamwise velocity difference $V_1-u_{\text{bg},1}(x_{\text{CM},2})$. Superimposed, we again plot the trajectories of the numerical Stokes flow simulations at higher $\textit{St}_\text{L}=\textit{St}_\text{rot.}=\textit{St}_\text{trans.}$. Here, we can again observe how the particle travels through regions of positive and negative forces, and that the velocity-oscillations will decrease in amplitude and approach a constant velocity as $\textit{St}_\text{L}$ is increased. 

To really conclude what is causing the drift, additional simulations were performed where $\textit{St}_\text{trans.}$ was varied independently from $\textit{St}_\text{rot.}$. The resulting drift velocities $V_\text{drift}$ after the initial transient are summarized in table~\ref{tab:TableStrotSttrans}. It is found that $V_\text{drift}$ is mainly dependent on the \emph{translational} inertia of the particle. Even though the oscillation period changes, the effect of $\textit{St}_\text{rot.}$ at a constant $\textit{St}_\text{trans.}$ is almost negligible. 

\begin{table}
\begin{center}
  \begin{tabular}{l l l l}
    {}&{$\textit{St}_\text{rot.}=1$}&{$\textit{St}_\text{rot.}=30$}&{$\textit{St}_\text{rot.}=300$}\\[3mm]
{$\textit{St}_\text{trans.}=1$}&{$1.1\times 10^{-4}$}&{$1.2\times 10^{-4}$}&{$1.8\times 10^{-4}$}\\[3mm]
{$\textit{St}_\text{trans.}=30$}&{$1.3\times 10^{-3}$}&{$1.4\times 10^{-3}$}&{$1.7\times 10^{-3}$}\\[3mm]
{$\textit{St}_\text{trans.}=300$}&{$2.8\times 10^{-4}$}&{$3.3\times 10^{-4}$}&{$2.5\times 10^{-4}$}\\[3mm]
  \end{tabular}
\caption{\label{tab:TableStrotSttrans}The final lateral velocity $V_\text{drift}$ for different combinations of $\textit{St}_\text{trans.}$ and $\textit{St}_\text{rot.}$ for a particle with $r_p=3$.}
\end{center}
\end{table}

It is also quite clear what will happen with oblate spheroids. Since the inertial oblate spheroid will drift towards a rotation around its symmetry axis, there will be no `jerk' in the translational motion, the force will be the same regardless of rotational phase and the particle will assume the velocity corresponding to zero force. There will thus not be any lateral motion for oblate particles.

\subsection{Additional remark}
The present results of the numerical Stokes flow simulations have demonstrated that the flow problem can be completely analyzed analytically by integrating the equations of motion \eqref{eq:GovEqMotion} using the force and torque expressions by \citet{Jeffery} and \citet{lamb1932hydrodynamics} and the modified expressions by \citet{Chwang} in \eqref{eq:LambTorque}--\eqref{eq:ChwangTorque} and \eqref{eq:LambForce}--\eqref{eq:ChwangForce}. The QBX method however can be utilized in the future for more complicated flow problems, e.g.~including complex geometries, where analytical solutions are difficult to obtain.

\section{Consequences for simulations with Lagrangian particles}\label{sec:LPTcons}

The fact that there is an additional force that arises from the second spatial derivative of the velocity also has consequences for Lagrangian particle tracking (LPT) methods. In these schemes, the force on the particle is calculated by knowing the (dimensional) velocity difference $\Delta\boldsymbol{U}^*=\boldsymbol{u}^*_\text{CM}-\boldsymbol{V}^*$ between the particle and the undisturbed fluid. Similarly, the torque is found only through the orientation and the local velocity gradients, i.e.~the first spatial derivatives of the velocity. The second spatial derivative was seen here to only affect the translation and not the rotation. So how can we evaluate if this contribution to the force is relevant?

Consider a particle in a Lagrangian frame centred on the particle with flow direction $\boldsymbol{e}_1$ and gradient direction $\boldsymbol{e}_2$ with constant curvature of the velocity profile given by $\dot{\gamma}'$. The particle experiences the velocity $\Delta\boldsymbol{U}^*$ and will thus have dimensional force contribution according to \citet{lamb1932hydrodynamics} as
\begin{equation}
\boldsymbol{F}^*_{\Delta U}=16\upi \mu l \mathsfbi{K}(\phi)\Delta\boldsymbol{U}^*.
\end{equation}
Since the Lagrangian frame is always centred on the particle, the force contribution due to the second spatial derivative of the velocity will be equivalent to setting $x_{\text{CM},2}=0$ in \eqref{eq:ChwangForce}. The dimensional result thus becomes
\begin{eqnarray}
\boldsymbol{F}^*_\text{curv.}&=&16\upi \mu l \dot{\gamma}' l^2 \frac{1-E^2\cos^2\phi}{3}\left(\cos\phi\frac{\boldsymbol{e}'_1}{2A_1}-\sin\phi\frac{\boldsymbol{e}'_2}{A_2}\right)\nonumber\\
&=&16\upi \mu l \dot{\gamma}' l^2 \frac{1-E^2\cos^2\phi}{3}\mathsfbi{K}(\phi)\boldsymbol{e}_1,
\end{eqnarray}
where $A_1=(-2E + (1+E^2) \log[(1+E)/(1-E)])/E^3$ and $A_2=(2E + (3E^2-1) \log[(1+E)/(1-E)])/E^3$ are parameters determined by the particle geometry. Furthermore, all geometry dependent parameters $\mathsfbi{K}, E, A_1$ and $A_2$ are $\textit{O}(1)$ for moderate aspect ratios. In order to neglect the contribution of the velocity curvature in an LPT simulation when calculating the force, the condition $|\dot{\gamma}'|l^2/|\Delta \boldsymbol{U}^*|\ll1$ should thus hold.

As a final remark, it is likely that there is a similar correction to the rotation by taking into account the third spatial derivative of the velocity.

\section{Comparison with gravitational forces}\label{sec:GravForceEffects}

\begin{figure}
\begin{center}
\includegraphics{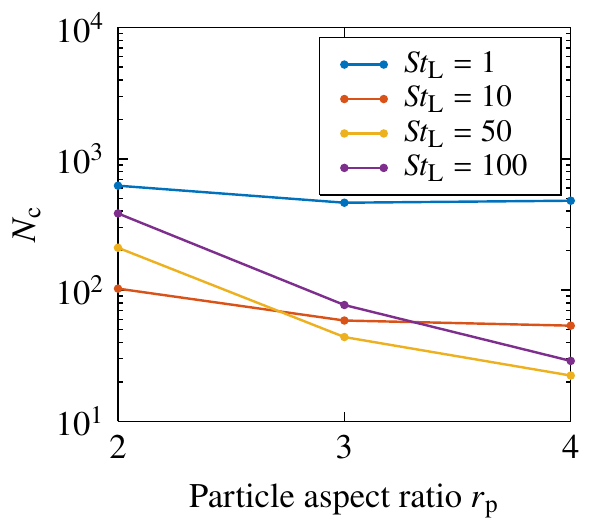}
\caption{\label{fig:CriticalNumber}Critical inertial drift number $N_\text{c}$ as function of $r_\text{p}$ and $\textit{St}_\text{L}$.}
\end{center}
\end{figure}

In this work we have observed that prolate spheroidal particles in a parabolic flow gives rise to a lateral drift. This is however only true as long as fluid inertia is neglected while particle inertia is dominating, i.e~$\textit{St}_\text{L}\gg \Rey_\text{p,L}$ and $\Rey_\text{p,L}\ll 1$, which means that the density ratio $\alpha=\rho_\text{p}/\rho_\text{f}$ must fulfill
\begin{equation}
\alpha\gg 1
\end{equation}
and
\begin{equation}
\alpha\gg \textit{St}_\text{L}.
\end{equation}
This new mechanism then seems to be relevant for solid particles suspended in air. However, as soon as $\rho_\text{p}\neq\rho_\text{f}$, there is sedimentation in the presence of gravity. We must therefore carefully examine when the dimensional drift velocity found here $V^{*}_\text{drift}=V_\text{drift}\dot{\gamma}'l^2$ ($V_\text{drift}$ is the dimensionless velocity found in fig.~\ref{fig:rpDep} depending on $r_\text{p}$ and $\textit{St}$) is larger than the sedimentation velocity $V^*_\text{sed}$. From \citet{KochShaqfeh}, the sedimentation velocity for prolate spheroids in Stokes flow can be expressed as
\begin{equation}
V^*_\text{sed}=\frac{1}{12}\frac{\rho_\text{f}(\alpha-1) gl^2}{\mu r_\text{p}^2}\beta(r_\text{p},\psi)\approx\frac{1}{12}\frac{\rho_\text{p} gl^2}{\mu r_\text{p}^2}\beta(r_\text{p},\psi)
\end{equation}
with the correction term $\beta(r_\text{p},\psi)$ being given by
\begin{equation}
\beta(r_\text{p},\psi)=b_1(r_\text{p})+b_2(r_\text{p})\cos \psi,
\end{equation}
where $\psi$ is the angle between gravity and particle orientation and the functions $b_1$ and $b_2$ are defined as
\begin{eqnarray}
  b_1(r_\text{p})=&\dfrac{r_\text{p}^2}{r_\text{p}^2-1}+\dfrac{2r_\text{p}^3-3r_\text{p}}{(r_\text{p}^2-1)^{3/2}}\ln\left[r_\text{p}+(r_\text{p}^2-1)^{1/2}\right],\\
  b_2(r_\text{p})=&\dfrac{-3r_\text{p}^2}{r_\text{p}^2-1}+\dfrac{2r_\text{p}}{(r_\text{p}^2-1)^{3/2}}\ln\left[r_\text{p}+(r_\text{p}^2-1)^{1/2}\right]\\
&+\dfrac{r_\text{p}-2r_\text{p}^3}{(r_\text{p}^2-1)^{3/2}}\ln\left[r_\text{p}-(r_\text{p}^2-1)^{1/2}\right].\nonumber
\end{eqnarray}
To neglect any effect of gravity we must consequently fulfill
\begin{subeqnarray}
V^{*}_\text{drift}&\gg& V^*_\text{sed,max},\\
  \text{i.e.}\qquad V_\text{drift}(r_\text{p},\textit{St})\dot{\gamma}'l^2&\gg&\frac{1}{12}\frac{\rho_\text{p} gl^2}{\mu r_\text{p}^2}\beta(r_\text{p},0),\\
  \label{eq:gravityCondition}
  \text{i.e.}\qquad \frac{\mu\dot{\gamma}'}{\rho_\text{p} g}&\gg&\frac{\beta(r_\text{p},0)}{12r_\text{p}^2V_\text{drift}(r_\text{p},\textit{St}_\text{L})}.
\end{subeqnarray}
We define the inertial drift number $N$ and its critical value $N_\text{c}$ as
\begin{eqnarray}
N&=&\frac{\mu\dot{\gamma}'}{\rho_\text{p} g},\\
N_\text{c}&=&\frac{\beta(r_\text{p},0)}{12r_\text{p}^2V_\text{drift}(r_\text{p},\textit{St}_\text{L})},
\end{eqnarray}
and rewrite the condition \eqref{eq:gravityCondition} as
\begin{equation}
N\gg N_\text{c}.
\end{equation}
In figure~\ref{fig:CriticalNumber} we plot the critical inertial drift number $N_\text{c}$ as a function of $r_\text{p}$ and find that $N_\text{c}=\textit{O}(10^2)$ for the moderate aspect ratios studied here. However, $N_\text{c}$ is much smaller for more slender spheroids at $\textit{St}_\text{L}=\textit{O}(10)$ and can possibly reach below $N_\text{c}=\textit{O}(10)$. 

For argument's sake, let us consider a slender fiber and that we need to fulfill  
\begin{equation}
N=\frac{\mu\dot{\gamma}'}{\rho_\text{p} g}\gg 10.
\end{equation}
Given that the solid particle is suspended in air somewhere on earth, we can assume that $\mu=10^{-5} \, \text{Pa s}$, $\rho_\text{f}=10^0 \, \text{kg/m$^3$}$, $\rho_\text{p}=10^3 \, \text{kg/m$^3$}$ and $g=10^1 \, \text{m/s$^2$}$. This gives us a condition on the velocity profile curvature of
\begin{equation}
\dot{\gamma}'\gg 10^{10} \, \text{m$^{-1}$s$^{-1}$}.
\end{equation}
How much is this? If we superimpose a constant flow in opposite direction, we will end up with a parabolic velocity profile similar to that in a planar channel flow. Given a plane Poiseuille flow, $\dot{\gamma}'$ can be expressed in terms of the pressure gradient
\begin{equation}
  \dot{\gamma}'=\frac{1}{\mu}\frac{\mathrm{d}p^*}{\mathrm{d}x^*},
\end{equation}
which means that in order to achieve the previous inequality, we need a pressure gradient of
\begin{equation}
  \frac{\mathrm{d}p^*}{\mathrm{d}x^*}\gg 10^5 \, \text{Pa/m},
\end{equation}
i.e.~a pressure drop of 1 bar per meter of the channel. In order to keep such a flow laminar given a channel of height $H$, the channel Reynolds number $\Rey$ must fulfill
\begin{subeqnarray}
\Rey=\frac{\rho_\text{f}\dot{\gamma}'H^3}{\mu}&\ll&10^3,\\
  \text{i.e.}\qquad H&\ll&10^{-4} \, \text{m}.
\end{subeqnarray}
For the particle to fit inside such a channel, it must be $l\ll 10^{-4} \, \text{m}$. Given our assumption of the particle and fluid density, the creeping flow assumption now reads
\begin{subeqnarray}
\kappa&\gg&\textit{St}_\text{L},\\
  \text{i.e.}\qquad \kappa&\gg&\frac{\rho_\text{p}\dot{\gamma}'l^3}{\mu},\\
  \text{i.e.}\qquad 10^3 \, \text{m$^{3}$} &\gg&\frac{10^3 \times 10^{10} l^3}{10^{-5}},\\
  \text{i.e.}\qquad l&\ll&10^{-5/3} \, \text{m},
\end{subeqnarray}
i.e.~the particle must be smaller than 2~mm, which is less strict than the previous assumption of the particle size. One might argue that the effects presented in this work then is dominating for slender fibers with the size of tenths of microns suspended in air. However, in any practical application, having these microfibers suspended in a channel of sub-millimeter size with a velocity around $U^*\approx 100 \, \text{m/s}$, lacks true relevance. Furthermore, we have not began to comment on other effects that might arise due to these higher pressures and small length dimensions of the particle, such as compressibility, temperatures, Brownian motion and the break-down of the continuum assumption of the fluid. These arguments quickly diverge into absurdity, and it is safe to assume that the inertial drift due to the quadratic velocity profile will never be dominating over sedimentation in any earthly application. 

Even if it is not dominating, the inertial migration described in this work should be regarded as a correction factor to a sedimentation process in a flow where fluid inertia of the flow around the particles is negligible but particle inertia is dominating due to a high solid-to-fluid density ratio.

\section{Conclusions}\label{sec:Conclusions}

In this work we have discussed the influence of a parabolic velocity profile on the dynamics of an inertial prolate spheroidal particle both in terms of translation and rotation. The results are derived with the assumption that inertia of the fluid flow around the particle is negligible. The work can thus be viewed as a direct continuation of the work by \citet{Chwang} to account for particle inertia.

It is found that the non-sphericity, particle inertia and the quadratic profile contribute to a lateral drift towards regions of higher shear. The particle inertial effects are governed by the local Stokes number $\textit{St}_\text{L}$ based on the local shear rate. In the absence of inertia ($\textit{St}_\text{L}=0$), when the particle is aligned in the flow-gradient plane, it assumes an intermittent tumbling motion where the angular velocity depends on the angle. This intermittent rotation is determined only by the local shear rate and described in detail by \citet{Jeffery}. This causes the particle in the parabolic profile to have an intermittent streamwise velocity depending on the instantaneous orientation \citep{Chwang}. 

With higher $\textit{St}_\text{L}$, the particle behaves exactly as expected in a linear shear flow according to the results by \citet{LundellCarlsson}. However, due to its translational inertia, the particle is not quick enough to adapt to the non-zero forces during a particle rotation. The final trajectory of the particle, when the impulse during a rotational period is zero, describes a translational motion with a lateral drift. At very high $\textit{St}_\text{L}$, the particle will approach a rotation with constant angular velocity and also a translation with constant velocity, and the lateral drift vanishes.

If the particle is not initially oriented in the flow-gradient plane it will drift towards such an orbit. Since the particle is seen to have angular dynamics which are identical to a particle in simple shear, we can know from \citet{LundellCarlsson} that the maximum orbit drift occurs at $\textit{St}_\text{L}\approx \textit{St}_{0.5}$. The critical value of $\textit{St}_{0.5}$ is defined when particle inertia is overcoming viscous damping and can be found analytically, see \eqref{eq:St05}. We find furthermore that this critical number also predicts when we can expect the maximum lateral drift. Since the particle tends to migrate towards regions of larger shear, we know that a particle released at a position where $\textit{St}_\text{L}<\textit{St}_{0.5}$ will reach both a position where $\textit{St}_\text{L}\approx \textit{St}_{0.5}$ with a maximum drift angle (angle between particle velocity and flow direction) and later have a decreasing drift angle as the particle migrates further in the lateral direction. 

The maximum lateral drift velocity was found to be dependent on the particle aspect ratio, with higher drift for larger aspect ratio $r_\text{p}$. However, since the intermittent tumbling motion is essential to have this type of drift, inertial oblate particles and spheres ($r_\text{p}\leq1$) are not affected by this drift since they preferably rotate in a non-intermittent state.

Although the results were found through simulations of the Stokes flow using the numerical QBX method, it is found that the forces and torques on the particle is exactly represented by the analytical expressions by \citet{lamb1932hydrodynamics} and the modified expressions of \citet{Chwang} (with modifications presented in appendix~\ref{appB}). This means that the equations of translational and rotational motion can be directly integrated using these expressions to arrive at the same conclusions as in the present work. The strength in the QBX method lies mainly in the fact that we can additionally use the same method for studying other flow problems where analytical solutions are more difficult to obtain, for example having multiple particles and wall-bounded flow through complex geometries.

In Lagrangian particle tracking (LPT) methods, the force on the particle is usually calculated using only the instantaneous velocity difference $\Delta\boldsymbol{U}^*$ between the particle and the undisturbed fluid. Here, it is demonstrated that there is also a contribution to the force from the curvature of the velocity profile $\dot{\gamma}'$, and that $|\dot{\gamma}'|l^2/|\Delta \boldsymbol{U}^*|\ll1$ must hold for this contribution to be negligble.

Finally, it is known from recent work \citep[e.g.~][]{Einarsson1,Einarsson2} that fluid inertial effects are dominating over particle inertial effects for spheroidal particles of same density as the surrounding fluid. The consequence is that we can only neglect fluid inertia, if the local Stokes number $\textit{St}_\text{L}$ is much larger than the local Reynolds number $\Rey_\text{p,L}\ll 1$, i.e.~the solid-to-fluid density ratio must be large. In the presence of gravity this analysis will however break down due to sedimentation if the particles are not density matched. We find that the inertial drift described in this work will never be the dominating cause for migration of heavy particles in a channel flow on earth, but will rather enter as a correction factor to a sedimentation process.



\backsection[Funding]{J.B. and A.-K.T. acknowledge financial support from the G\"{o}ran Gustafsson Foundation for Research in Natural Sciences and Medicine. T.R. and F.L. acknowledge financial support from Wallenberg Wood Science Center (WWSC), Bengt Ingestr\"{o}m's Foundation and \AA Forsk (\AA ngpannef\"{o}reningen’s Foundation for Research and Development).}

\backsection[Declaration of interests]{The authors report no conflict of interest.}




\appendix

\section{QBX parameters}\label{appA}

The parameters used are given in table~\ref{tab:TableQBX}. Here, $n_\phi$
and $n_\theta$ are the number of grid points on the spheroid in
the periodic and non-periodic direction, respectively (as
explained in Section~\ref{sec:DiscQuad}); $r/h$ controls the
distance from each expansion centre to the particle surface
($h=2\upi l r_\text{p}^{-1}/n_\phi$); $\kappa$ is the upsampling
factor of the grid; $p$ denotes the order of the expansions. The
Matlab functions \texttt{gmres} and \texttt{ode113} are used for
solving systems of linear equations and ordinary differential equations, respectively. For a more detailed explanation of these parameters, we refer to \citet{Klinteberg16}.

\begin{table}
\begin{center}
  \begin{tabular}{r@{$\:=\:$}l}
    $n_{\phi}$ & 16 \\
    $n_{\theta}$ & $r_\text{p} n_{\phi}$ \\
    $r/h$ & 0.6 \\
    $\kappa$ & 30 \\
    $p$ & 30 \\
    Tolerance for \verb|gmres| & $10^{-12}$ \\
    Tolerance for \verb|ode113| & $10^{-8}$ \\
  \end{tabular}
  \caption{\label{tab:TableQBX}Parameters used for the QBX method.}
\end{center}
\end{table}

\section{Derivation of the force in a parabolic velocity profile}\label{appB}

\citet{Chwang} uses a paraboloidal flow $\boldsymbol{u}_\text{bg}^\text{Chw}(\boldsymbol{x}) = (1/2) (x_2^2+x_3^2) \boldsymbol{e}_1$, which leads to the force $\boldsymbol{F}^\text{Chw} = \boldsymbol{F}_0 + \boldsymbol{F}^\text{par}_\text{Chw}$, where $\boldsymbol{F}_0$ is as in \eqref{eq:LambForce} and
\begin{eqnarray}
  \label{eq:ChwangCorrection}
  \boldsymbol{F}^\text{par}_\text{Chw}(\phi, \boldsymbol{x}_\text{CM}) &=& \frac{8 \upi E^3 \cos\phi}{3} \frac{3x_{\text{CM},2}^2 + \textcolor{blue}{2-E^2} - E^2 \cos^2\phi}{-2E + (1+E^2) \log[(1+E)/(1-E)]} \boldsymbol{e}'_1 + \\
  && - \frac{16 \upi E^3 \sin\phi}{3} \frac{3x_{\text{CM},2}^2 + \textcolor{blue}{2-E^2} - E^2 \cos^2\phi}{2E + (3E^2 - 1) \log[(1+E)/(1-E)]} \boldsymbol{e}'_2.\nonumber
\end{eqnarray}
By adjusting the derivation by \citet{Chwang} to the parabolic flow $\boldsymbol{u}_\text{bg}(\boldsymbol{x}) = (1/2) x_2^2 \boldsymbol{e}_1$ used in this paper, we find that the force is $\boldsymbol{F} = \boldsymbol{F}_0 + \boldsymbol{F}^\text{par}$ as in Section~\ref{sec:TranslationDiscussion}. The terms that differ between $\boldsymbol{F}^\text{par}_\text{Chw}$ and $\boldsymbol{F}^\text{par}$ are highlighted blue in \eqref{eq:ChwangCorrection}; cf.~\eqref{eq:ChwangForce}. The torque on the particle is the same in the parabolic flow $\boldsymbol{u}_\text{bg}(\boldsymbol{x})$ and the paraboloidal flow $\boldsymbol{u}_\text{bg}^\text{Chw}(\boldsymbol{x})$.

\citet{Chwang} found that a particle with $\textit{St}_\text{trans.}=0$ moving in the paraboloidal flow $\boldsymbol{u}_\text{bg}^\text{Chw}(\boldsymbol{x})$ will have the velocity
\begin{equation}
  \boldsymbol{V}^\text{Chw}(\phi,\boldsymbol{x}_\text{CM})  = \frac{1}{6} \left( 3x_{\text{CM},2}^2 +  \textcolor{blue}{2-E^2} - E^2\cos^2\phi \right) \boldsymbol{e}_1.
\end{equation}
In the parabolic flow $\boldsymbol{u}_\text{bg}{\boldsymbol{x}}$ the velocity $\boldsymbol{V}$ is instead given by \eqref{eq:VelocityByChwang}. Note that the difference
\[
  \boldsymbol{V}^\text{Chw} - \boldsymbol{V} = \frac{1}{6} (1-E^2) \boldsymbol{e}_1
\]
is independent of $\phi$ and $\boldsymbol{x}_\text{CM}$.

\bibliographystyle{jfm}
\bibliography{references}

\end{document}